%% file: sample-sigconf.tex
\documentclass[sigconf]{acmart}
\makeatletter
\def\@ACM@checkaffil{
    \if@ACM@instpresent\else
    \ClassWarningNoLine{\@classname}{No institution present for an affiliation}%
    \fi
    \if@ACM@citypresent\else
    \ClassWarningNoLine{\@classname}{No city present for an affiliation}%
    \fi
    \if@ACM@countrypresent\else
        \ClassWarningNoLine{\@classname}{No country present for an affiliation}%
    \fi
}
\makeatother
\AtBeginDocument{%
  }

\setcopyright{acmlicensed}
\copyrightyear{2018}
\acmYear{2018}
\acmDOI{XXXXXXX.XXXXXXX}

\acmConference[DOSSA '24]{}{June 30,
  2024}{Buenos Aires, Argentina}
\acmISBN{978-1-4503-XXXX-X/18/06}

\setcopyright{none}
\renewcommand\footnotetextcopyrightpermission[1]{}
\begin{document}

\title{Towards Generalized On-Chip Communication for Programmable Accelerators in Heterogeneous Architectures}

\author{ \parbox{\linewidth}{\centering Joseph Zuckerman$^1$, John-David Wellman$^2$,  Ajay Vanamali$^1$, Manish Shankar$^1$, Gabriele Tombesi$^1$, Karthik Swaminathan$^2$, Kevin Lee$^1$, Mohit Kapur$^2$, Robert Philhower$^2$, Pradip Bose$^2$, Luca P. Carloni$^1$
}}
\affiliation{%
  \institution{Columbia University, Department of Computer Science$^1$; IBM Thomas J. Watson Research Center$^2$
  }
}
\renewcommand{\shortauthors}{J. Zuckerman et al.}

\begin{abstract}
We present several enhancements to the open-source ESP platform to support flexible and efficient on-chip communication for programmable accelerators in heterogeneous SoCs. These enhancements include 1) a flexible point-to-point communication mechanism between accelerators, 2) a multicast NoC that supports data forwarding to multiple accelerators simultaneously, 3) accelerator synchronization leveraging the SoC's coherence protocol, 4) an accelerator interface that offers fine-grained control over the communication mode used, and 5) an example ISA extension to support our enhancements. Our solution adds negligible area to the SoC architecture and requires minimal changes to the accelerators themselves. We have validated most of these features in complex FPGA prototypes and plan to include them in the open-source release of ESP in the coming months.
\end{abstract}

\maketitle
\sloppy
\input{sections/1-intro}

\vspace{-0.5cm}
\input{sections/2-esp}
\input{sections/3-enhancements}
\input{sections/4-results}
\input{sections/5-future}

\begin{acks}
    We thank Paolo Mantovani for providing the baseline implementation of the NoC router and for his advice in implementing multicast. This work was supported in part by a National Science Foundation Graduate Research Fellowship. The views, opinions and/or other findings expressed are those of the authors and should not be interpreted as representing the official views or policies (either expressed or implied) of the National Science Foundation or the U.S Government.
\end{acks}

\bibliographystyle{ieeetr}
\bibliography{ref}
\end{document}

%% file: sections/1-intro.tex
\section{Introduction}

\textit{Programmable accelerators}~\cite{libra, plasticine, transmuter} have become a prominent element of system-on-chip (SoC) architectures thanks to their ability to balance energy-efficient and high-performance computation with flexibility for application developers. In contrast to their fixed-function counterparts, programmable accelerators execute instructions generated by a compiler, a specialized tool, or written by hand.
In some cases, they can be programmed using a \textit{domain-specific language}~\cite{dsagen_isca20}, a specialized language for a particular class of applications.  

A heterogeneous SoC architecture \cite{smiv,gonzalez_esscirc21,jia_esscirc22,gao_decades23, dossantos_isscc24} might feature several instances of programmable accelerators alongside general-purpose cores, fixed-function accelerators, and various peripherals. In various domains, workloads can be partitioned across several accelerators to exploit parallelism. There may also be data dependencies across kernels running on different accelerators; this can require synchronization among accelerators for correctness and direct forwarding of data for efficiency. Software developers writing applications for these complex systems would therefore benefit from a flexible on-chip communication substrate that supports multiple types of data transfer modes and seamless synchronization. 

\begin{figure}[!t]
    \centering
    \includegraphics[width=\linewidth]{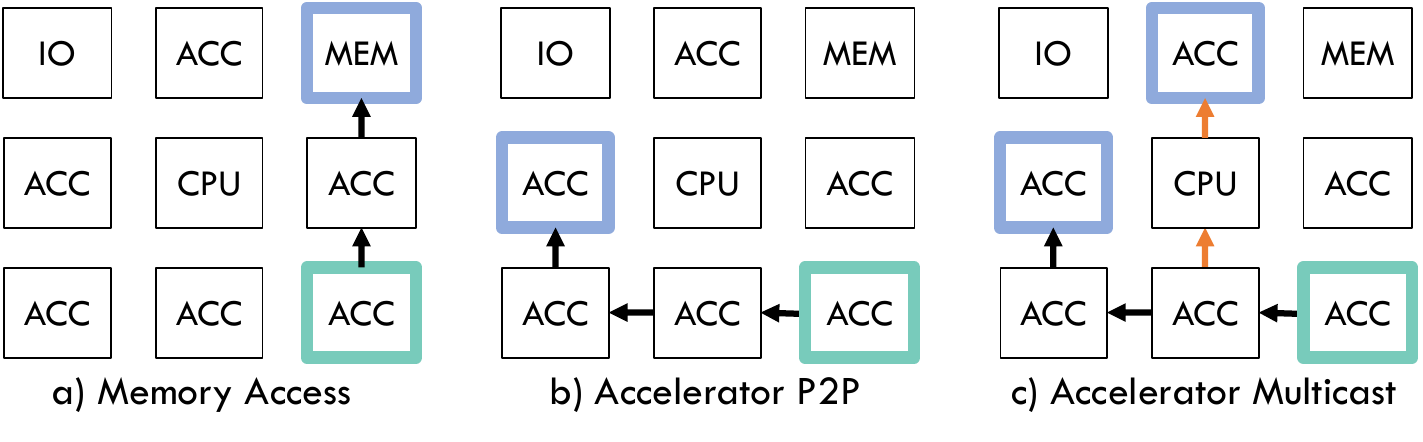}
    \vspace{-0.5cm}
    \caption{Three distinct data access modes for an accelerator in a 3x3 tile heterogeneous SoC.}
    \label{fig:access_modes}
    \vspace{-0.5cm}
\end{figure}

\figurename~\ref{fig:access_modes} shows 3 different data-access patterns that might be required by an accelerator in a heterogeneous SoC. In this case, the figure shows a 3x3 tile SoC, with 6 accelerators, 1 CPU, 1 memory tile, and 1 tile for IO peripherals. Typically, a \textit{network-on-chip} (NoC), serves as the interconnect in such a tiled system. The data access modes shown include: 1) direct memory access (DMA), which could be to a scratchpad, a last-level cache partition, or off-chip DRAM; 2) direct point-to-point (P2P) communication between 2 accelerators, in which outputs of one accelerator are directly forwarded as inputs to another; and 3) multicast transfer, in which outputs of one accelerator are directly forwarded to multiple others. Moreover, these different types of communication modes might be required within a single accelerator \textit{invocation}, which is the typical granularity of synchronization with a host. So, software-based solutions would require costly synchronization overheads. 

Instead, we propose hardware-based solutions, tightly integrated into the system-level architecture of a heterogeneous SoC, to support the desired communication and synchronization primitives. Our implementation requires minimal area overhead on top of an existing SoC architecture and only minor changes to the design of accelerators themselves. We build our solution on top of ESP~\cite{mantovani_iccad20}, an open-source platform for SoC design, but the main principles can apply to other SoC architectures. Our solution builds on the following contributions: \\
    \textbullet~Enhancements to ESP's existing P2P capabilities for flexibility. \\
    \textbullet~A lightweight and efficient multicast NoC, that integrates with ESP'S P2P capabilities. \\
    \textbullet~A proposal for inter-accelerator synchronization based on coherence provided by the SoC architecture. \\
    \textbullet~An accelerator interface that supports fine-grained control over communication modes and integrates well with existing standards. \\
    \textbullet~Example ISA extensions for programmable accelerators to leverage our architecture. \\

%% file: sections/2-esp.tex
\section{The ESP Accelerator Socket}

\begin{figure}[!t]
    \centering
    \includegraphics[width=0.9\linewidth]{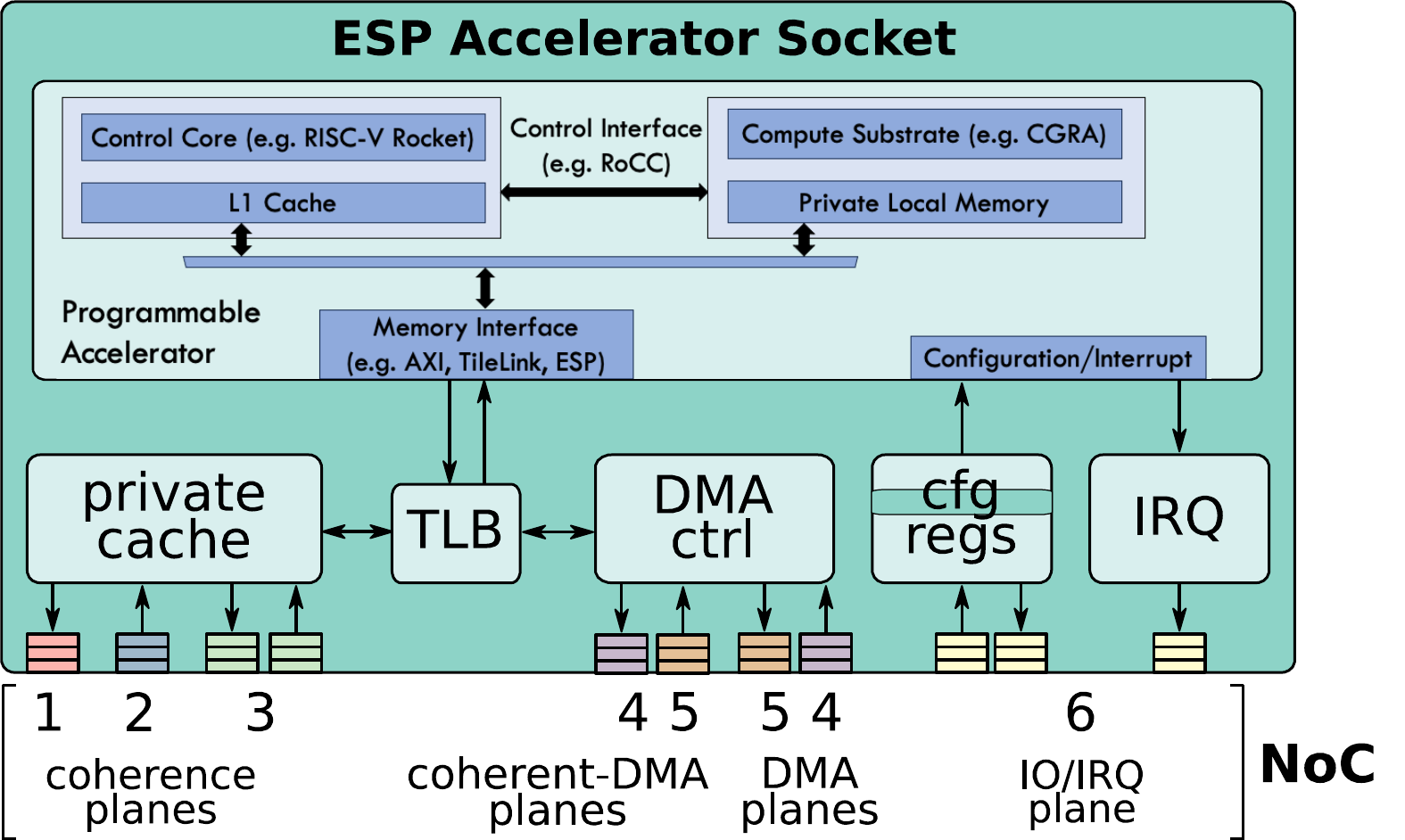}
    \caption{The ESP accelerator socket with an instance of a programmable accelerator.}
    \vspace{-0.5cm}
    \label{fig:socket}
\end{figure}

The ESP architecture is a heterogeneous tile grid, connected by a 2D mesh NoC. \figurename~\ref{fig:socket} shows one of the key types of tiles, the \textit{accelerator tile}, with an instance of an example programmable accelerator. Programmable accelerators need to feature dedicated structures for instruction dispatch, scheduling, and retirement~\cite{dyser}. To avoid defining an ISA and developing control structures from scratch, open-source ISAs like RISC-V and accompanying core implementations, such as the Rocket Core \cite{asanovic_rocket}, can be attractive for those developing programmable accelerators~\cite{overgen_micro2022}. In the case of the Rocket Core, ISA extensions can be used to communicate with a custom datapath through its RoCC interface. This datapath and custom private local memory (PLM) are the key to providing speedup for the target applications.

In ESP, accelerators are \textit{loosely coupled}~\cite{cota_dac15}, which means they are decoupled from the implementation of host cores, are connected through the system interconnect, and execute coarse-grained tasks on large workloads. Therefore, programmable accelerators in such architectures also need a memory interface to load large quantities of data from the rest of the system.

The dark green part of \figurename~\ref{fig:socket} shows the ESP \textit{accelerator socket}~\cite{carloni_dac16}. The socket decouples the design of the accelerator and provides various \textit{platform services}, such as DMA, address translation, interrupts, and configuration registers that greatly simplify the integration of the accelerator into the SoC. In ESP, accelerators are allocated a single contiguous, virtual buffer, which is potentially scattered across multiple large, physical pages. The TLB in the socket translates the accelerator virtual address to a global physical address. The data can then be accessed with one or more \textit{coherence modes}~\cite{giri_nocs18}. In \textit{DMA-based modes}, requests and responses use the 2 DMA NoC planes to send requests either to the last-level cache or directly to off-chip memory. If the accelerator socket instantiates an optional L2 cache, the \textit{fully-coherent mode} can be used to have the accelerator participate in the system's coherence protocol, in this case MESI, through the 3 coherence NoC planes.

ESP's P2P communication~\cite{giri_date20} reuses the DMA NoC planes to forward data directly from one accelerator to another. P2P communication adopts a \textit{pull-based mechanism} in order to satisfy the \textit{consumption assumption}, in which messages put on the NoC are always consumed~\cite{song_tpds03}, thereby preventing deadlock. The consumer accelerator initiates the transaction by sending a P2P request to the producer. If configured in the P2P mode, the producer waits for the request from the consumer before forwarding the data. 

%% file: sections/3-enhancements.tex
\section{Enhancements to ESP}
In this section, we detail the various enhancements made to the ESP architecture to support our proposal.

\textbf{Flexible P2P Communication.} The current implementation of ESP's P2P communication comes with a few restrictions. First, the selected communication mode for the accelerator must be applied to all transactions for the entire invocation (i.e. all memory or all P2P). In many cases, however, it would be convenient to access some data from memory and some data from another accelerator (e.g. a neural-network accelerator fetching model parameters from memory and a previous layer's outputs from another accelerator). We modified the accelerator socket to be able to change its communication mode at the granularity of each data transfer (i.e. burst); this is described further in the \textit{Accelerator Interface} section.

Second, the existing implementation of P2P requires that both the producer and consumer accelerator have the same data access pattern (i.e. number and size of bursts). This limits the applicability of P2P, particularly between accelerators of different types. By adding a length to each P2P request made by the consumer, this constraint is now relaxed, and the two accelerators can leverage P2P with different access patterns, only subject to the constraint that they must produce/consume the same total amount of data for a P2P transaction.

\textbf{Multicast NoC.} The NoC forms the backbone of the ESP architecture and has several key properties that are important for the functionality and performance of ESP. For example, because we leverage multiple physical planes instead of virtual channels and employ lookahead routing, there is a single cycle latency from router to router; moreover, the use of dimension-ordered routing guarantees the absence of routing deadlock. In designing our multicast NoC, we sought to minimize the changes to the ESP NoC. Starting from the design of the existing ESP NoC router, we modified the metadata of each NoC message to include a list of destinations instead of a single one. This metadata is contained in a NoC message's \textit{header flit}, the first unit of data exchanged between 2 routers; the number of supported destinations is therefore dependent on the bitwidth of the NoC. The lookahead routing logic, which computes the next routing instruction for a destination, is replicated to compute the direction for each destination in parallel. We modified the router to be able to forward a packet to multiple output ports in parallel when necessary, e.g. the bottom middle ACC tile in \figurename~\ref{fig:access_modes}c.

\begin{figure}[!t]
    \centering
    \includegraphics[width=\linewidth]{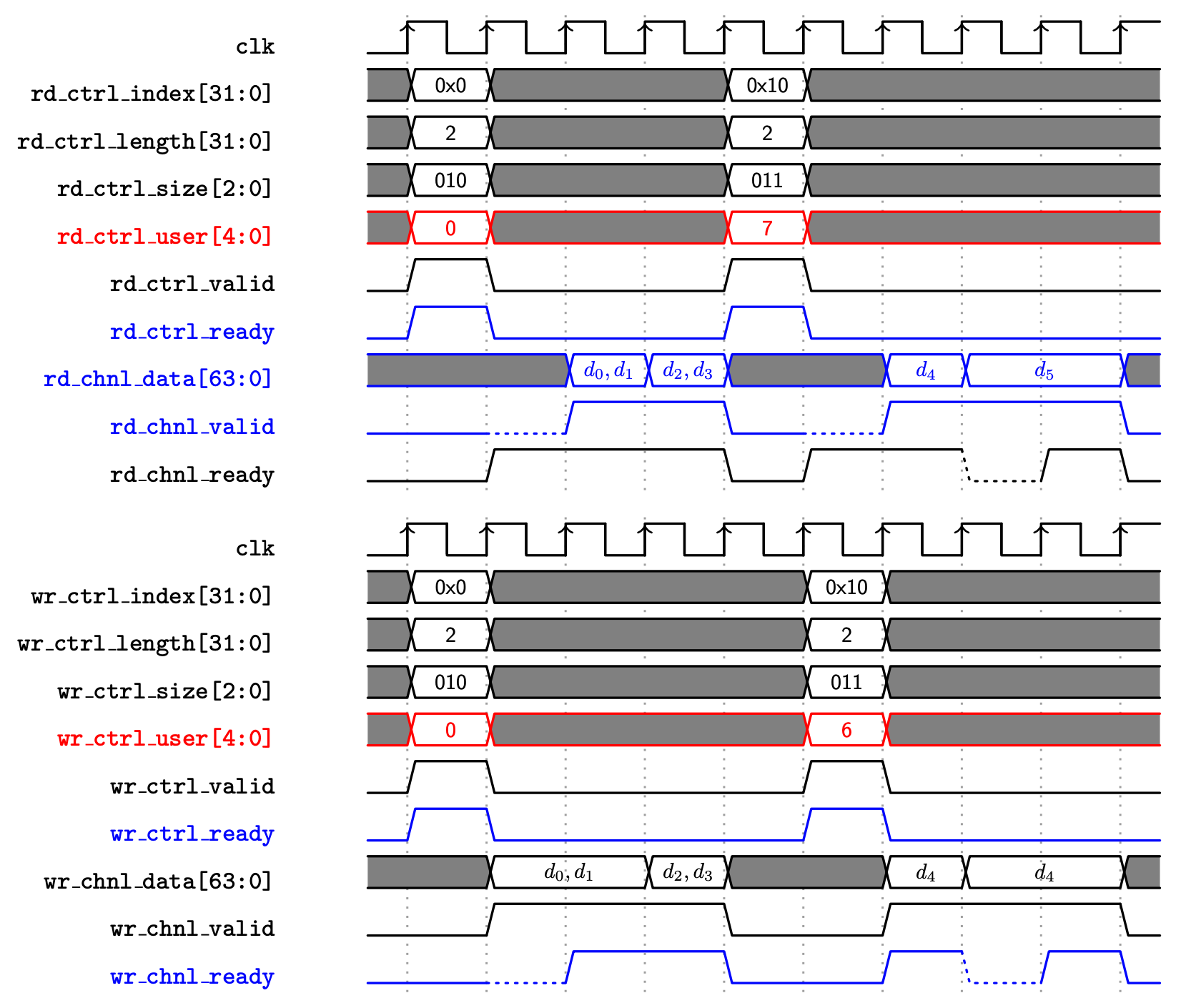}
    \caption{Signals of the 4 latency-insensitive channels of the ESP accelerator interface.}
    \vspace{-0.5cm}
    \label{fig:acc_interface}
\end{figure}

We integrated the new multicast NoC with ESP's P2P capabilities to facilitate multicast transfers between accelerators. When a multicast transfer is specified, the producer uses the P2P exchange mode, but instead of waiting for a single consumer request, it waits for the number of consumers specified. Once all requests have been received, the producer creates the appropriate multicast message with the coordinates of all destinations encoded and then sends the data to all of them in a single transfer.

\textbf{Accelerator Synchronization.} Rather than designing a bespoke synchronization solution for a particular type of accelerator, we propose a more general-purpose synchronization scheme leveraging the ESP coherence protocol. As previously mentioned, ESP can optionally instantiate an L2 cache in the accelerator tile, which enables the accelerator's participation in the MESI coherence protocol. However, the fully-coherent mode can be much less efficient than DMA modes for many workloads~\cite{giri_ieeemicro18, zuckerman_micro21}. Furthermore, similar to the current restrictions on switching between P2P and memory access, the same coherence mode must be applied for all data transfers within a single accelerator invocation. Therefore, we plan to reserve some portion of the accelerator's dataset for synchronization messages, which leverage fully-coherent transfers, while all other bulk transfers can leverage the DMA controller; some modest changes to the socket are required to support this.

\textbf{Accelerator Interface.} \figurename~\ref{fig:acc_interface} shows waveforms characterizing the updated ESP accelerator interface. Black and blue represent the existing signals coming from the accelerator and socket, respectively; new signals coming from the accelerator are in red.
The interface consists of 4 independent \textit{latency-insensitive}~\cite{carloni_tcad01_lip} channels: read control, read data, write control, and write data.

Each control channel contains signals to specify the length, word size, and address (relative to the accelerator's virtual buffer). The data channels merely carry the read and write data. We added a \texttt{user} field to each control channel to support our changes for flexible transfers and multicast. On the read channel, the \texttt{user} field encodes the \textit{source} of each transaction. Zero encodes a standard DMA request, while 1 to $(N - 1)$ encode a P2P request to one of the other accelerators in the SoC. A small, configurable lookup table in the socket encodes the tile coordinates for each index, so that these values can be \textit{virtualized}.
On the write channel, the \texttt{user} field encodes the number of destinations for the write: zero again encodes a DMA request, 1 encodes a unicast P2P transfer, and 2 to $(N-1)$ encode a multicast transfer.


Although this proposed interface builds on the ESP accelerator interface, it could be applied to other standards, in particular AXI~\cite{arm_ace}, which also has independent, latency-insensitive channels that serve similar purposes. 

\textbf{Example ISA.} We propose a simple, 2-instruction extension to the accelerator's ISA to govern DMA transactions: \textit{Initiate DMA request} (IDMA) and \textit{Check DMA} (CDMA). The IDMA instruction specifies the necessary information for the read/write control interfaces, including the length, word size, and source/number of destinations. It also specifies the virtual address (which is mapped to the global address space) to read from/\textit{write to} and the local physical address to store the read data to/\textit{fetch the write data from} in the accelerator's PLM. The IDMA instruction also returns a \textit{tag}, which uniquely identifies the DMA transaction locally to the accelerator. Because the DMA transactions are performed asynchronously with respect to the accelerator's pipeline, the CDMA instruction can then use the tag issued from IDMA to query the status of a particular DMA operation.  The CDMA instruction returns status information, which can be used by the accelerator for subsequent control flow, e.g. the accelerator can initiate a DMA to load data, do some computation, and then query whether the DMA load is complete, at which point the accelerator can proceed to compute with that data.

%% file: sections/4-results.tex
\section{Results}

While these enhancements to ESP greatly improve the flexibility of on-chip communication, particularly for programmable accelerators, the main quantitative results of our work come from the implementation and integration of the multicast NoC in ESP. In particular, in this section we detail the area overhead of adding multicast support to the NoC router and the speedups provided by leveraging multicast on a many-accelerator SoC prototyped on FPGA.

We first synthesized various configurations of the NoC router by sweeping both its bitwidth and the maximum number of supported multicast destinations. Because the header flit is used to encode the coordinates of each multicast destinations, the number of possible destinations is limited by the bitwidth of the NoC. For example, a 64-bit NoC can encode up to 5 destinations, and a 128-bit NoC can encode up to 14 destinations. In the current implementation, ESP supports multicasts of up to 16 destinations, but this could be expanded in the future. 

\begin{figure}
    \centering
    \includegraphics[width=\linewidth]{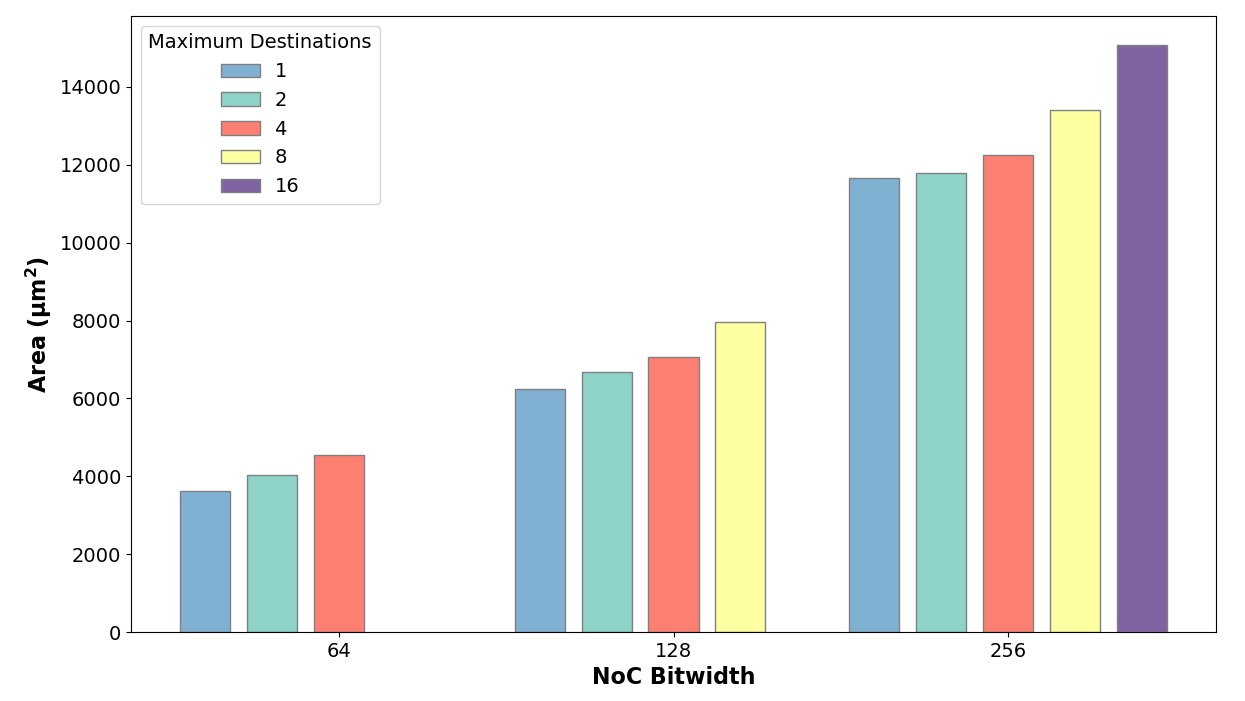}
    \caption{Area of a single NoC router with different bitwidths and maximum multicast destinations.}
    \label{fig:router_area}
\end{figure}

We synthesized the NoC with Cadence Genus targeting a 12$nm$ technology. \figurename~\ref{fig:router_area} shows the post-synthesis area of each router configuration. The baseline router (64 bits, no multicast) has an area of 3620$\mu m^2$. Increasing the bitwidth of the NoC shows a roughly proportional increase in the area of the router; this is expected, as much of the router area is occupied by the input queues. The 128-bit NoC and 256-bit NoC without multicast have areas of 6,230$\mu m^2$ and 11,520$\mu m^2$, respectively. Supporting additional multicast destinations comes at a cost of 200$\mu m^2$, on average, which is 5.5\%, 3.2\%, and 1.7\% of the 64-bit, 128-bit, and 256-bit baseline routers, respectively. The 64-bit, 128-bit, and 256-bit NoC routers can support 4, 8, and 16 destinations, respectively, with less than a 30\% increase of area. In summary, adding multicast incurs modest area overheads.

\begin{figure}
    \centering
    \includegraphics[width=0.7\linewidth]{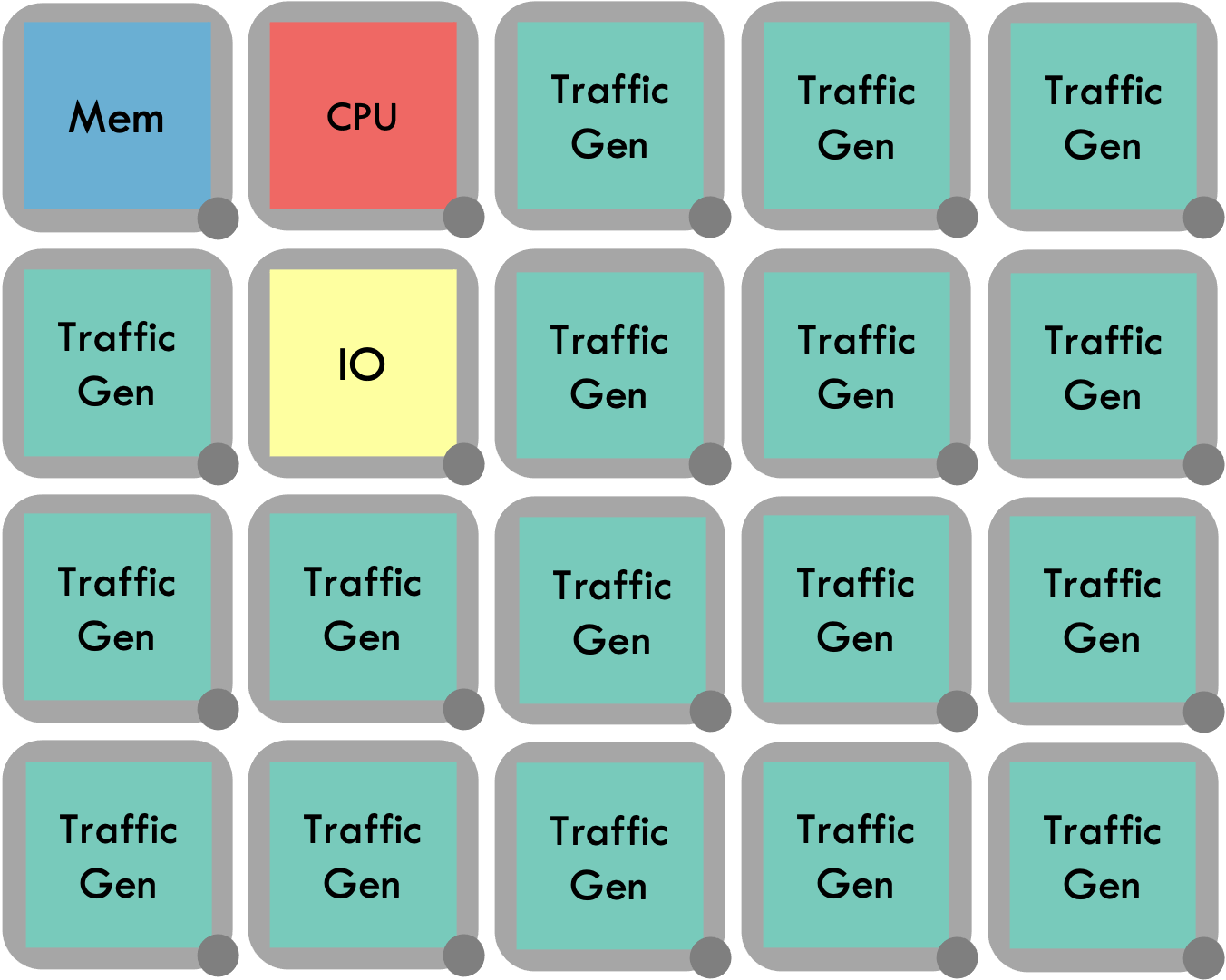}
    \caption{Evaluated 3x4 SoC with 1 CPU tile, 1 Memory tile, 1 IO tile, and 17 traffic generator accelerators.}
    \label{fig:mcast_soc}
\end{figure}

Next, we evaluated the performance benefits of leveraging multicast by running a toy application on an FPGA prototype of many-accelerator SoC. \figurename~\ref{fig:mcast_soc} shows the layout of the target SoC. It is a 12-tile SoC arranged in a 3x4 2D mesh with 1 CPU tile featuring the RISC-V CVA6 core \cite{ariane_paper, zuckerman_carrv22}, 1 Memory tile, 1 I/O tile, and 17 traffic generator accelerators. The traffic generator is used to mimic the communication patterns of an accelerator in the SoC, but does not perform any computation. In particular, our traffic generator accelerator performs the identity function, i.e. it writes the same data as output that it receives as input. We leverage a 256-bit NoC for communication between accelerators, which allows us to test
multicast up to the maximum of 16 destinations. Our SoC is implemented for a Xilinx Virtex Ultrascale+ VCU128 board and the design runs at 78 MHz. 

Our application mimics a dataflow of 1 producer accelerator that creates data that is used by $N$ consumer accelerators. We compare using multicast to a baseline of communication through shared memory (i.e. the producer writes to main memory and then the $N$ consumers read the same data). We vary both the number of consumer accelerators and the amount of data exchanged between accelerators. The traffic generator accelerator is capable of loading 4KB of data at a time; hence, larger data set sizes require multiple read and write bursts.

\begin{figure}
    \centering
    \includegraphics[width=\linewidth]{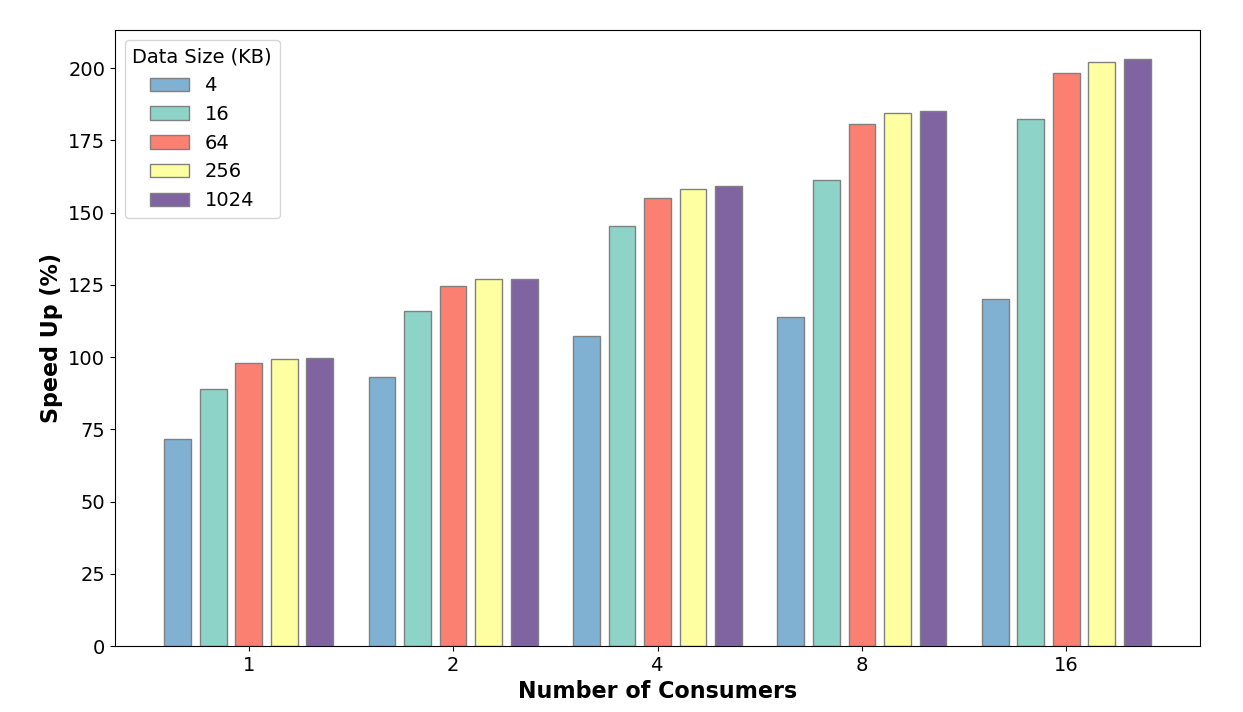}
    \caption{Speedup of multicast compared to shared-memory baseline with varying consumers and data size.}
    \label{fig:mcast_perf}
\end{figure}

\figurename~\ref{fig:mcast_perf} shows the speedup of multicast compared to the shared-memory baseline for each configuration of number of consumers and data size. Even with only 1 consumer (i.e. no multicast) and the smallest data set, we see a 72\% speedup compared to the baseline. Using P2P communication avoids a round trip to main memory and allows for finer-grained synchronization and pipelining across accelerators. As expected, adding additional consumers improves the speedup; with the same dataset size, a multicast to 16 consumers gives a speedup of 120\%. For $N$ consumers, we do not see a speedup of a factor of $N$, because we are not turning a purely serial operation (although there is a memory bottleneck in the baseline, there is some overlap in the execution of the accelerators) into a purely parallel one (the multicast has synchronization overheads that require some degree of serialization). As the dataset sizes increases, the speedup improves because the multicast P2P communication allows for the pipelining of the execution of the producer and consumers at the granularity of bursts, thereby hiding memory access latency and invocation overheads. This phenomenon plateaus at 1MB, when these overheads become negligible compared to the total size of the task. A maximum speedup of 203\% is achieved with 16 consumers and a 1MB workload.

%% file: sections/5-future.tex
\section{Conclusion and Future Work}
We presented a proposal for a system-level architecture that supports flexible and efficient on-chip communication for programmable accelerators in heterogeneous SoC architectures. We have completed the design of the flexible P2P, multicast NoC, and updated accelerator interface. The accelerator synchronization
is under development. Because of the substantial time required to design a programmable accelerator, the completed features have been validated on complex FPGA prototypes using traffic-generator accelerators and show significant performance improvements with modest overheads. We leave evaluation with real programmable accelerators, which we are actively developing, for a future paper. The completed features are already available publicly in development branches of ESP's GitHub. All of these features will eventually become part of the main public release~\cite{esp}.

%% file: sample-sigconf.bbl
\begin{thebibliography}{10}

\bibitem{libra}
Y.~Park, J.~J.~K. Park, H.~Park, and S.~Mahlke, ``{Libra: Tailoring SIMD Execution Using Heterogeneous Hardware and Dynamic Configurability},'' in {\em {2012 45th Annual IEEE/ACM International Symposium on Microarchitecture}}, pp.~84--95, 2012.

\bibitem{plasticine}
R.~Prabhakar, Y.~Zhang, D.~Koeplinger, M.~Feldman, T.~Zhao, S.~Hadjis, A.~Pedram, C.~Kozyrakis, and K.~Olukotun, ``{Plasticine: A reconfigurable architecture for parallel patterns},'' in {\em {2017 ACM/IEEE 44th Annual International Symposium on Computer Architecture (ISCA)}}, pp.~389--402, 2017.

\bibitem{transmuter}
S.~Pal, S.~Feng, D.-h. Park, S.~Kim, A.~Amarnath, C.-S. Yang, X.~He, J.~Beaumont, K.~May, Y.~Xiong, {\em et~al.}, ``{Transmuter: Bridging the efficiency gap using memory and dataflow reconfiguration},'' in {\em {Proceedings of the ACM International Conference on Parallel Architectures and Compilation Techniques}}, pp.~175--190, 2020.

\bibitem{dsagen_isca20}
J.~Weng, S.~Liu, V.~Dadu, Z.~Wang, P.~Shah, and T.~Nowatzki, ``Dsagen: Synthesizing programmable spatial accelerators,'' in {\em 2020 ACM/IEEE 47th Annual International Symposium on Computer Architecture (ISCA)}, pp.~268--281, 2020.

\bibitem{smiv}
S.~K. Lee, P.~N. Whatmough, M.~Donato, G.~G. Ko, D.~Brooks, and G.-Y. Wei, ``Smiv: A 16-nm 25-mm² soc for iot with arm cortex-a53, efpga, and coherent accelerators,'' {\em IEEE Journal of Solid-State Circuits}, vol.~57, no.~2, pp.~639--650, 2022.

\bibitem{gonzalez_esscirc21}
A.~Gonzalez, J.~Zhao, B.~Korpan, H.~Genc, C.~Schmidt, J.~Wright, A.~Biswas, A.~Amid, F.~Sheikh, A.~Sorokin, S.~Kale, M.~Yalamanchi, R.~Yarlagadda, M.~Flannigan, L.~Abramowitz, E.~Alon, Y.~S. Shao, K.~Asanović, and B.~Nikolić, ``A 16mm2 106.1 gops/w heterogeneous risc-v multi-core multi-accelerator soc in low-power 22nm finfet,'' 2021.

\bibitem{jia_esscirc22}
T.~Jia, P.~Mantovani, M.~C. dos Santos, D.~Giri, J.~Zuckerman, E.~J. Loscalzo, M.~Cochet, K.~Swaminathan, G.~Tombesi, J.~J. Zhang, N.~Chandramoorthy, J.-D. Wellman, K.~Tien, L.~Carloni, K.~Shepard, D.~Brooks, G.-Y. Wei, and P.~Bose, ``{A 12nm Agile-Designed SoC for Swarm-Based Perception with Heterogeneous IP Blocks, a Reconfigurable Memory Hierarchy, and an 800MHz Multi-Plane NoC},'' in {\em European Solid-State Circuits Conference (ESSCIRC)}, September 2022.

\bibitem{gao_decades23}
F.~Gao, T.-J. Chang, A.~Li, M.~Orenes-Vera, D.~Giri, P.~J. Jackson, A.~Ning, G.~Tziantzioulis, J.~Zuckerman, J.~Tu, K.~Xu, G.~Chirkov, G.~Tombesi, J.~Balkind, M.~Martonosi, L.~Carloni, and D.~Wentzlaff, ``Decades: A 67mm2, 1.46tops, 55 giga cache-coherent 64-bit risc-v instructions per second, heterogeneous manycore soc with 109 tiles including accelerators, intelligent storage, and efpga in 12nm finfet,'' in {\em 2023 IEEE Custom Integrated Circuits Conference (CICC)}, pp.~1--2, 2023.

\bibitem{dossantos_isscc24}
M.~Cassel~dos Santos, T.~Jia, J.~Zuckerman, M.~Cochet, D.~Giri, E.~J. Loscalzo, K.~Swaminathan, T.~Tambe, J.~J. Zhang, A.~Buyuktosunoglu, K.-L. Chiu, G.~D. Guglielmo, P.~Mantovani, L.~Piccolboni, G.~Tombesi, D.~Trilla, J.-D. Wellman, E.-Y. Yang, A.~Amarnath, Y.~Jing, B.~Misra, J.~Park, V.~Suresh, S.~Adve, P.~Bose, D.~Brooks, L.~Carloni, K.~Shepard, and G.-Y. Wei, ``{A 12nm Linux-SMP-Capable RISC-V SoC with 14 Accelerator Types, Distributed Hardware Power Managment, and Flexible NoC-Based Data Orchestration},'' in {\em International Solid-State Circuits Conference (ISSCC)}, February 2024.

\bibitem{mantovani_iccad20}
P.~Mantovani, D.~Giri, G.~Di~Guglielmo, L.~Piccolboni, J.~Zuckerman, E.~G. Cota, M.~Petracca, C.~Pilato, and L.~P. Carloni, ``Agile soc development with open esp,'' in {\em 2020 IEEE/ACM International Conference On Computer Aided Design (ICCAD)}, pp.~1--9, IEEE, 2020.

\bibitem{dyser}
V.~Govindaraju, C.-H. Ho, and K.~Sankaralingam, ``{Dynamically Specialized Datapaths for energy efficient computing},'' in {\em {2011 IEEE 17th International Symposium on High Performance Computer Architecture}}, pp.~503--514, 2011.

\bibitem{asanovic_rocket}
K.~Asanovic, R.~Avizienis, J.~Bachrach, S.~Beamer, D.~Biancolin, C.~Celio, H.~Cook, D.~Dabbelt, J.~Hauser, A.~Izraelevitz, S.~Karandikar, B.~Keller, D.~Kim, J.~Koenig, Y.~Lee, E.~Love, M.~Maas, A.~Magyar, H.~Mao, M.~Moreto, A.~Ou, D.~A. Patterson, B.~Richards, C.~Schmidt, S.~Twigg, H.~Vo, and A.~Waterman, ``The {Rocket Chip} generator,'' Tech. Rep. UCB/EECS-2016-17, UC Berkeley, 2016.

\bibitem{overgen_micro2022}
S.~Liu, J.~Weng, D.~Kupsh, A.~Sohrabizadeh, Z.~Wang, L.~Guo, J.~Liu, M.~Zhulin, R.~Mani, L.~Zhang, J.~Cong, and T.~Nowatzki, ``Overgen: Improving fpga usability through domain-specific overlay generation,'' in {\em 2022 55th IEEE/ACM International Symposium on Microarchitecture (MICRO)}, pp.~35--56, 2022.

\bibitem{cota_dac15}
E.~G. Cota, P.~Mantovani, G.~D. Guglielmo, and L.~P. Carloni, ``An analysis of accelerator coupling in heterogeneous architectures,'' in {\em Proceedings of the ACM/IEEE Design Automation Conference (DAC)}, 2015.

\bibitem{carloni_dac16}
L.~P. Carloni, ``The case for {Embedded Scalable Platforms},'' in {\em Proceedings of the ACM/IEEE Design Automation Conference (DAC)}, pp.~17:1--17:6, 2016.

\bibitem{giri_nocs18}
D.~Giri, P.~Mantovani, and L.~P. Carloni, ``{NoC}-based support of heterogeneous cache-coherence models for accelerators,'' in {\em Proceedings of the International Symposium on Networks-on-Chip (NOCS)}, pp.~1:1--1:8, 2018.

\bibitem{giri_date20}
D.~Giri, K.-L. Chiu, G.~D. Guglielmo, P.~Mantovani, and L.~P. Carloni, ``{ESP4ML: P}latform-based design of systems-on-chip for embedded machine learning,'' in {\em Proceedings of the IEEE Conference on Design, Automation, and Test in Europe (DATE)}, 2020.

\bibitem{song_tpds03}
Y.~H. Song and T.~Pinkston, ``A progressive approach to handling message-dependent deadlock in parallel computer systems,'' {\em IEEE Transactions on Parallel and Distributed Systems}, vol.~14, no.~3, pp.~259--275, 2003.

\bibitem{giri_ieeemicro18}
D.~Giri, P.~Mantovani, and L.~P. Carloni, ``Accelerators {\&} coherence: An {SoC} perspective,'' {\em IEEE Micro}, vol.~38, no.~6, pp.~36--45, 2018.

\bibitem{zuckerman_micro21}
J.~Zuckerman, D.~Giri, J.~Kwon, P.~Mantovani, and L.~P. Carloni, ``{Cohmeleon: Learning-Based Orchestration of Accelerator Coherence in Heterogeneous SoCs},'' in {\em Proceedings of the IEEE/ACM Symposium on Microarchitecture (MICRO)}, 2021.

\bibitem{carloni_tcad01_lip}
L.~P. Carloni, K.~L. McMillan, and A.~L. Sangiovanni-Vincentelli, ``Theory of latency-insensitive design,'' {\em IEEE Transactions on CAD of Integrated Circuits and Systems}, vol.~20, no.~9, pp.~1059--1076, 2001.

\bibitem{arm_ace}
{ARM}, ``{AMBA AXI and ACE Protocol Specification}.'' \url{https://developer.arm.com/documentation/ihi0022/h}, 2020.

\bibitem{ariane_paper}
F.~Zaruba and L.~Benini, ``The cost of application-class processing: Energy and performance analysis of a {Linux}-ready {1.7-GHz 64-Bit RISC-V} core in {22-nm FDSOI} technology,'' {\em IEEE Transactions on Very Large Scale Integration Systems}, vol.~27, no.~11, pp.~2629--2640, 2019.

\bibitem{zuckerman_carrv22}
J.~Zuckerman, P.~Mantovani, D.~Giri, and L.~P. Carloni, ``{Enabling Heterogeneous, Multicore SoC Research with RISC-V and ESP},'' 2022.

\bibitem{esp}
{Columbia SLD Group}, ``{ESP Release}.'' {\scriptsize \url{www.esp.cs.columbia.edu}}, 2019.

\end{thebibliography}
